\documentclass[superscriptaddress]{elsarticle}
\usepackage{graphicx}
\usepackage{amsmath}
\usepackage{amssymb}
\usepackage{bm}
\usepackage{color}
\usepackage[english]{babel}

\makeatletter
\def\ps@pprintTitle{%
  \let\@oddhead\@empty
  \let\@evenhead\@empty
  \let\@oddfoot\@empty
  \let\@evenfoot\@oddfoot
}
\makeatother

\newcommand{\Fder}{\mathbf{F}}
\newcommand{\n}{\vec n}
\newcommand{\ee}{\vec e}
\newcommand{\B}{\mathbf{B}}
\newcommand{\C}{\mathbb{C}}
\newcommand{\Cc}{\mathbf{C}}
\newcommand{\K}{\mathbf{K}}
\newcommand{\Tt}{\mathbf{T}}
\newcommand{\D}{\vec D}
\newcommand{\E}{\vec E}
\newcommand{\I}{\mathbf{I}}
\newcommand{\Q}{\mathbf{Q}}
\newcommand{\A}{\mathbf{A}}
\newcommand{\R}{\mathbf{R}}
\newcommand{\M}{\mathcal{M}}
\newcommand{\ab}{\mathbf{a}}
\newcommand{\bb}{\mathbf{b}}
\newcommand{\Pp}{\mathbf{P}}
\newcommand{\Bnn}{B_{nn}}
\newcommand{\Btnn}{B_{nn}^{(2)}}
\newcommand{\Btrnn}{B_{nn}^{(3)}}
\newcommand{\Bdd}{B_{DD}}
\newcommand{\Btdd}{B_{DD}^{(2)}}
\newcommand{\Btrdd}{B_{DD}^{(3)}}
\newcommand{\Bnd}{B_{Dn}}
\newcommand{\Btnd}{B_{Dn}^{(2)}}
\newcommand{\Btrnd}{B_{Dn}^{(3)}}
\newcommand{\BnnInv}{B_{nn}^{(-1)}}
\newcommand{\BtnnInv}{B_{nn}^{(-2)}}

\newcommand{\BndInv}{B_{Dn}^{(-1)}}
\newcommand{\BtndInv}{B_{Dn}^{(-2)}}

\newcommand{\BddInv}{B_{DD}^{(-1)}}
\newcommand{\BtddInv}{B_{DD}^{(-2)}}
\newcommand{\BtrddInv}{B_{DD}^{(-3)}}

\begin{document}

\begin{frontmatter}
\title{Manipulating P-and S-elastic waves in dielectric elastomers via external electric stimuli}

\author{Pavel I. Galich}

\author{Stephan Rudykh}
\address{Department of Aerospace Engineering, Technion -- Israel Institute of Technology, Haifa 32000, Israel}
\ead{rudykh@technion.ac.il}

\begin{abstract}
We investigate elastic wave propagation in finitely deformed dielectric elastomers in the presence of an electrostatic field. To analyze the propagation of both longitudinal (P-) and transverse (S-) waves, we utilize \emph{compressible} material models. We derive
 explicit expressions of the generalized acoustic tensor and phase velocities of elastic waves for the ideal and enriched dielectric elastomer models. We analyze the slowness curves of the elastic wave propagation, and find the P-S-mode disentangling phenomenon. In particular, P- and S- waves are separated by the application of an electric field. The divergence angle between P- and S-waves strongly depends on the applied electrostatic excitation. The influence of the electric field is sensitive to material models. Thus, for ideal dielectric model the in-plane shear velocity increases with an increase in electric field, while for the enriched model the velocity may decreases depending on material constants. Similarly, the divergence angle gradually increases with an increase in electric field, while for the enriched model, the angle may be bounded. Material compressibility affects the P-wave velocity, and, for relatively compressible materials, the slowness curves evolve from circular to elliptical shapes manifesting in an increase of the reflection angle of P-waves. As a results, the divergence angle decreases with an increase in material compressibility.

\end{abstract}

\end{frontmatter}
%

%
%

\section{Introduction}

\begin{figure}[h]
 \centering{\includegraphics[scale=0.3]{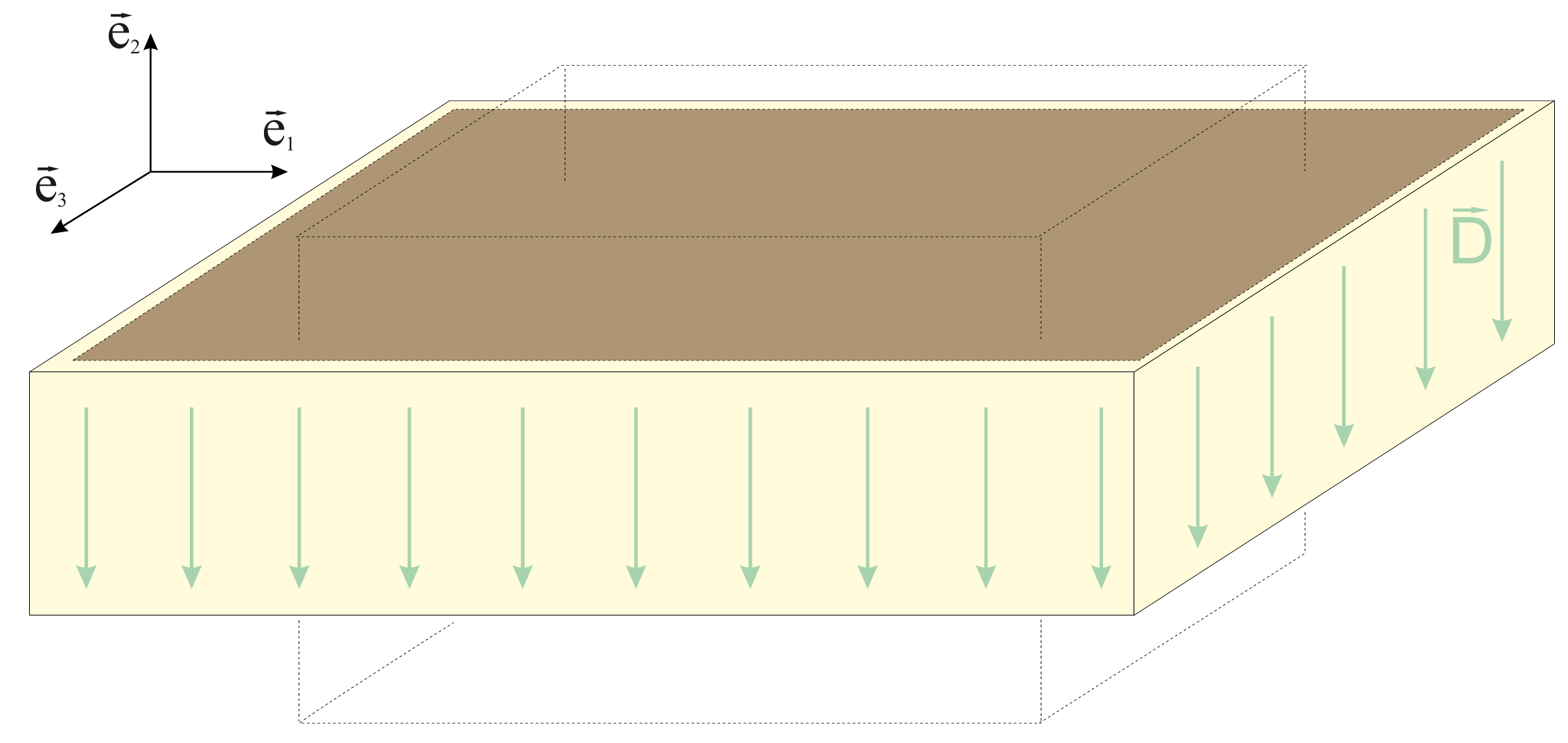}}
 \caption{\footnotesize DE layer subjected to an electric field.}\label{deformDE}
 \end{figure}

 Dielectric elastomers (DEs) are soft responsive materials that can change their form and shape  when subjected to electric stimuli~\cite{Pelrine&al2000,barcohen2002}.
 DEs have attracted considerable attention due to a large variety of possible applications ranging from artificial muscles and soft robotics to energy conversion and noise cancelling devises \cite{Halloran&al2008jap,brochu2010,carpi2011dielectric,rudy&etal12ijnm,kornbluh&al2012,rogers2013sciense,nguyen2014}.
   Figure~\ref{deformDE} schematically illustrates a typical set-up of a planar DE actuator with deformation induced by applied voltage \cite{wissler2005,Wissler&Mazza2007,hossain&etal15aam}. The DE layer coated with compliant electrodes contracts due to electrostatic attraction between oppositely charged electrodes.

 The theoretical framework of the non-linear electroelasticity is based on the theory first developed by Toupin~\cite{toupin1956,toupin1963}, which was recently revisited by Dorfmann and Ogden \cite{dorfmann2005,dorfmann2010} and Suo \cite{suo2008,Suo2010amss}. This followed by a series of works on modeling DEs \cite{thyl&etal12sms,volokh12jam,itskov&kheim14mms,keip&etal14cmame,cohen&debotton14ejma,jabareen15iutam,miehe&etal15ijnme,aboudi15ijss}, to name a few recent contributions. The electromechanical coupling in typical DEs is rather weak, and, therefore, DEs need to operate at the edge of instabilities and breakthrough voltages to achieve meaningful actuation \cite{PhysRevB.76.134113,plante2007,zhao&suo2008,zhao&suo2010prl,rudy&debo11zamp,rudykhetal2013,rudy&bertoldi13jmps,Gei&al2014ijss,siboni&etal14mms}. Potentially, the need in the high voltage can be reduced through architectured microstructures of DEs and increasing the electromechanical coupling \cite{Cheng&al2004apl,rudykh&etal2013,cao&zhao13apl,galipeau&etal14ijss}, another promising approach is synthesis of new soft dielectric materials \cite{madsen&etal14polymer}.

\begin{figure}[h]
         \centering{\includegraphics[scale=0.55]{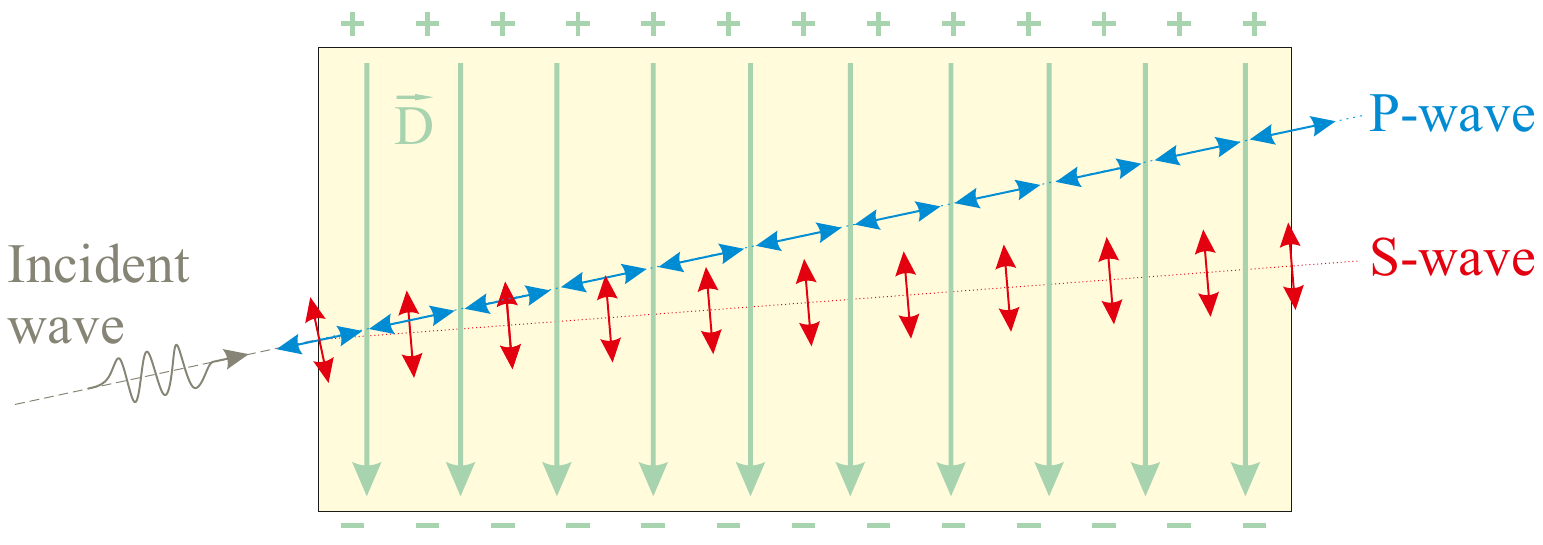}}
        \caption{\footnotesize Schematics of the splitting of P-and S-elastic waves in a nearly incompressible neo-Hookean DE. P-wave does not refract from the initial direction of propagation, while S-wave refracts for a certain angle.}\label{splNH}
      \end{figure}

The effective electromechanical properties of DEs can be actively controlled by external electric stimuli. This opens an opportunity to manipulate wave propagation in DEs by applying electric field. In this work we focus on elastic wave propagation in finitely deformed DEs in the presence of an electric field. It is known that elastic wave propagation in purely elastic materials depends on the media mechanical properties and on the deformation \cite{RudykhBoyce2014prl,galich&rudykh2015,galich&rudykh15comment}. In DEs the material moduli are modified by external electric field and induced deformations. To explore the elastic wave propagation in DEs, we follow the widely used approach of the analysis of small-amplitude motions superimposed on the finite deformations \cite{ogden97book} induced by an external field \cite{dorfmann2010}. To allow consideration of longitudinal wave propagation (as opposite to the recent works \cite{Gei&al2011,Shmuel&al2012ijnm,Chen&Dai2012amss}), we utilize {\it compressible} electroactive material models, namely the  ideal~\cite{PhysRevB.76.134113} and enriched DE models. By application of wave propagation analysis to the compressible DE models, we derive explicit expressions of generalized acoustic tensor and phase velocities for both longitudinal (P-) and transverse (S-) waves. We find that electrostatically induced changes in the important characteristics of shear and pressure waves lead to the disentangling phenomenon, where P- and S-waves travel in different directions. A schematic illustration of the P- and S-wave splitting phenomenon is shown in Fig.~\ref{splNH}.
 The divergence angle strongly depends on the external field, induced deformation, and material compressibility.
    P- and S-waves possess diverse properties and, thus, can be used for different purposes. For example, S-wave serves as a virtual "finger" to probe the elasticity of the internal regions of the body in the shear wave elasticity imaging~\cite{sarvazyan1998}. Longitudinal and transversal elastic waves can be separated on interfaces between dissimilar materials; however, in this case loss of energy or/and wave-mode conversion is encountered on the interface~\cite{achenbach1973}. On this account we can use the phenomenon to split P-and S-elastic waves in dielectric elastomers by application of a bias electric field. This feature can be used, for example, in small-scale micro-electromechanical systems, where it is convenient to use electric field to control the performance.
%

\section{Analysis}
To analyse the finitely deformed state, we introduce the deformation gradient $\Fder (\vec{X},t)=\nabla_{\vec{X}}\otimes\vec{x}(\vec X ,t)$, where $\vec X$ and $\vec x$ are position vectors in the reference and current configurations, respectively. In order to model a non-linear electroelastic material behaviour, we consider an energy function $\psi(\Fder,\D_0)$ that is a function of the deformation gradient $\Fder$ and electric displacement vector $\D_0$ represented in the reference configuration. The corresponding electric displacement in the current configuration is given by $\D=J^{-1}\Fder\cdot\D_0$, where $J=\text{det}~\Fder$. Thus, the first Piola-Kirchhoff stress tensor and electric field in the reference configuration are given by
\begin{equation}\label{Piola&EF}
\Pp_0=\frac{\partial \psi}{\partial \Fder} \quad \text{and}\quad \E_0=\frac{\partial \psi}{\partial \D_0},
\end{equation}
corresponding counterparts in the current configuration are
\begin{equation}
\Tt=J^{-1}\Pp_0\cdot\Fder^T \quad \text{and}\quad \E=\Fder^{-T}\cdot\E_0.
\end{equation}
 Next, we consider the small-amplitude motions superimposed on finite deformations. Incremental forms of the constitutive equations~\eqref{Piola&EF} are
\begin{equation}\label{inc_rel0}
\begin{split}
\delta\Pp_0=\C_0:\delta\Fder+\M_0\cdot\delta\D_0,\\
 \delta\E_0=\delta\Fder:\M_0+\K_0\cdot\delta\D_0,
\end{split}
\end{equation}
where $\delta$ denotes an increment change in a value on the right side; $\C_0,\M_0$ and $\K_0$ are the so-called tensors of electroelastic moduli~\cite{dorfmann2010} defined as
\begin{equation}\label{el_t_ref}
 \C_0=\frac{\partial^2\psi}{\partial \Fder \partial \Fder},\M_0=\frac{\partial^2\psi}{\partial \Fder \partial \D_0}, \K_0=\frac{\partial^2\psi}{\partial \D_0 \partial \D_0}.
\end{equation}
Recall that $\C_0=\C_0^{(3412)}$ and $\K_0=\K_0^T$, where superscript (3412) denotes an isomer of the fourth-rank tensor~\cite{ryzhak93jmps,nikitin&ryzhak2008,galich&rudykh2015} as detailed in \ref{Appendix A}.

The incremental constitutive equations~\eqref{inc_rel0} in the frame of the updated reference configuration are
\begin{equation}\label{inc_up_conf}
\delta\Pp=\C:\delta\mathbf{H}+\M\cdot\delta\D_{01}, \delta\E_{01}=\delta\mathbf{H}:\M+\K\cdot\D_{01},
\end{equation}
where $\delta\Pp=J^{-1}\delta\Pp_0\cdot\Fder^T, \delta\D_{01}=J^{-1}\Fder\cdot\delta\D_0, \delta\E_{01}=\Fder^{-T}\cdot\delta\E_0,\delta\mathbf{H}=\delta\Fder\cdot\Fder^{-1}$ and
\begin{equation}\label{el_t_cur}
\begin{split}
\C=&\ J^{-1}\left(\Fder\cdot\C_0^{(2134)}\cdot\Fder^T\right)^{(2134)},\\
\M=&\ \Fder\cdot\M_0^{(213)}\cdot\Fder^{-1},\\
\K=&\ J\Fder^{-T}\cdot\K_0\cdot\Fder^{-1}.
\end{split}
\end{equation}
Note that $\C=\C^{(3412)}, \M^{(213)}=\M$ and $\K=\K^T$.
The linearised equation of motion and Maxwell's equations in Eulerian form are
\begin{equation}\label{eqm&ME}
\nabla_{\vec{x}}\cdot\delta\Pp=\rho\frac{\partial^2 \vec{u}}{\partial t^2},
\nabla_{\vec{x}}\times\delta\E_{01}=0, \nabla_{\vec{x}}\cdot\delta\D_{01}=0,
\end{equation}
where $\vec{u}$ is incremental displacement and $\rho$ is material density.
We seek for a solution of equations~\eqref{eqm&ME} in the form of plane waves with constant polarization
\begin{equation}\label{sol}
\vec{u}=\vec{m}f(\n\cdot\vec{x}-ct),\D_{01}=\vec{d}g(\n\cdot\vec{x}-ct),
\end{equation}
where $f$ is a twice continuously differentiable function and $g$ is a continuously differentiable function; unit vectors $\vec{m}$ and $\vec{d}$ are polarization vectors of mechanical and electrical displacement, respectively; the unit vector $\n$  defines the direction of propagation of the wave, and $c$ is phase velocity of the wave.

By substitution~\eqref{inc_up_conf} and~\eqref{sol} into~\eqref{eqm&ME}, we obtain
\begin{equation}
\mathbf{A}\cdot\vec{m}=\rho c^2 \vec{m},
\end{equation}
where $\mathbf{A}$ is the so-called "generalized" acoustic tensor which defines the condition of propagation of plane elastic waves in non-linear electroelastic materials. Moreover, for electroelastic materials with an arbitrary strain energy function  $\psi(\Fder,\D_0)$ it has the following form~\cite{Lopez2015}
\begin{equation}\label{AcLP}
 \mathbf{A}=\mathbf{Q}-\frac{2}{(\text{tr} \hat\K)^2-\text{tr} \hat\K^2}\mathbf{R}\cdot\left((\text{tr} \hat\K)\hat\I -\hat \K\right)\cdot\mathbf{R}^T,
\end{equation}
where
\begin{equation}\label{Pn}
\hat \I =\I-\n\otimes\n
\end{equation}
 is the projection on the plane normal to $\n$, $\hat\K=\hat \I \cdot \K \cdot \hat \I$ and
\begin{equation}\label{Qq}
\Q=\C^{(1324)}:\n\otimes\n, \R=\n\cdot\M.
\end{equation}
It is worth mentioning that the generalized acoustic tensor $\A$ is symmetric as well as the purely elastic "classic" acoustic tensor $\mathbf{Q}$. Recall that for DE to be stable, generalized acoustic tensor $\A$ has to be positively defined~\cite{Lopez2015}. Detailed descriptions of the non-linear electroelastic theory for DEs can be found in works of Dorfmann and Ogden~\cite{dorfmann2005,dorfmann2010}, and Suo~\cite{suo2008,Suo2010amss}.

It is well known that incompressible materials do not support longitudinal waves. Therefore, to analyse both transversal and longitudinal waves, we consider  the energy function $\psi(\Fder,\D_0)$ for the {\it compressible} electroelastic material in the following form:
\begin{equation}\label{genEF}
\psi(\Fder,\D_0)=\psi_{elas}(\Fder)+\frac{1}{2\epsilon J}\left(\gamma_0 I_{4e} + \gamma_1 I_{5e} + \gamma_2 I_{6e} \right),
\end{equation}
where $\psi_{elas}(\Fder)$ is the purely elastic energy function (for example, neo-Hookean, Mooney-Rivlin, Gent, etc.~\cite{ogden97book}), $\epsilon$ is the material permittivity in the undeformed state $\left(\Fder=\I\right)$ and $\gamma_i$ are dimensionless parameters, moreover $\gamma_0+\gamma_1+\gamma_2=1$, and
\begin{equation}
I_{4e}=\D_0\cdot\D_0, I_{5e}=\D_0\cdot\Cc\cdot\D_0, I_{6e}=\D_0\cdot\Cc^2\cdot\D_0
\end{equation}
are invariants depending on the electric displacement $\D_0$ and right Cauchy-Green tensor $\Cc=\Fder^T\cdot\Fder$.

For energy function~\eqref{genEF} the relation between the electric displacement and electric field is
\begin{equation}
\E=\frac{1}{\epsilon}\left(\gamma_0 \B^{-1} + \gamma_1 \I + \gamma_2 \B \right)\cdot\D,
\end{equation}
where $\B=\Fder\cdot\Fder^T$ is the left Cauchy-Green tensor.
The corresponding tensors of electroelastic moduli are
\begin{equation}\label{EMT}
\begin{split}
\C=&\C_{elas}+\frac{1}{2\epsilon}\left(\gamma_0 \C_{4e} + \gamma_1 \C_{5e} + \gamma_2 \C_{6e} \right),\\
\M=&\frac{1}{2\epsilon}\left(\gamma_0 \M_{4e} + \gamma_1 \M_{5e} + \gamma_2 \M_{6e} \right),\\
\K=&\frac{1}{\epsilon}\left(\gamma_0 \B^{-1} + \gamma_1 \I + \gamma_2 \B \right),
\end{split}
\end{equation}
where $\C_{elas}$ is derived from the purely elastic part $\psi_{elas}(\Fder)$ of the energy function in accordance to \eqref{el_t_ref} and \eqref{el_t_cur}; and the explicit relations for the other tensors are given in the~\ref{Appendix B}.

Finally, the corresponding generalized acoustic tensor takes the following form
\begin{equation}\label{GA1}
\A=\Q_{elas}+\frac{1}{2\epsilon}\left(\gamma_0 \Q_{4e} + \gamma_1 \Q_{5e} + \gamma_2 \Q_{6e} -\frac{4}{\eta}\A_e\right),
\end{equation}
where $\Q_{elas}$ is calculated by applying of Eq. \eqref{el_t_ref},  \eqref{el_t_cur}  and \eqref{Qq} to the 
purely elastic part  $\psi_{elas}(\Fder)$ of the energy function; the explicit expressions for $\eta$, and tensors $\Q_{4e}$, $\Q_{5e}$, $\Q_{6e}$ and $\A_e$ are given in the~\ref{Appendix B}. 
Remarkably, in the particular case of $\Fder=\lambda_1\ee_1\otimes\ee_1+\lambda_2\ee_2\otimes\ee_2+\lambda_3\ee_3\otimes\ee_3$, $\n=\ee_1$ and $\D_0=D_2\ee_2$ generalized acoustic tensor~\eqref{GA1} reduces to
\begin{equation}
\A=\Q_{elas}+\frac{\gamma_2 D_2^2}{\epsilon \lambda_3^2}\ee_2\otimes\ee_2,
\end{equation}
where ($\ee_1,\ee_2,\ee_3$) is the orthonormal basis.

\begin{figure}[h]
  \centering{\includegraphics[scale=0.7]{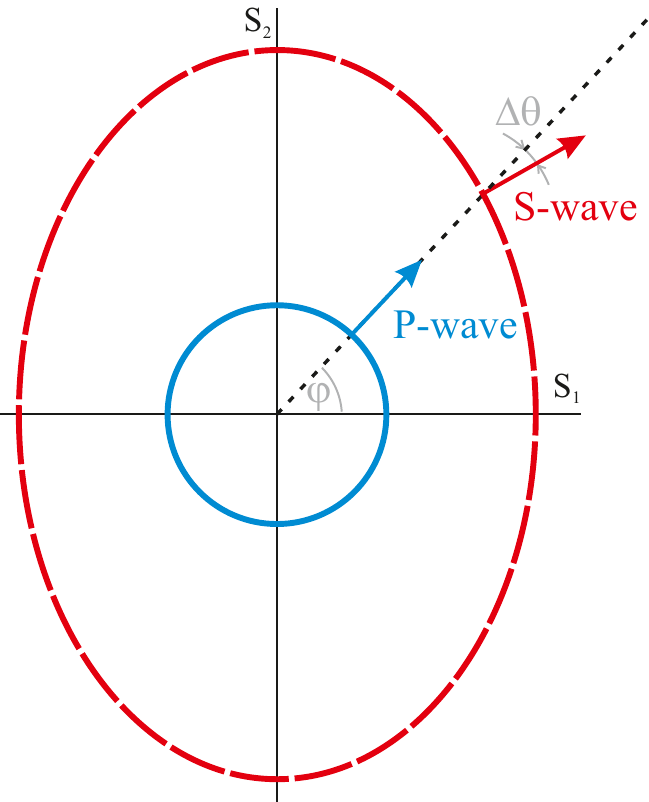}}
 \caption{\footnotesize Slowness curves of pressure and shear waves in the nearly incompressible DE subjected to a voltage (for clarity, slowness curve of P-wave is zoomed in).}\label{slowness}
\end{figure}

Let us consider how the application of an electric field affects the P-and S-wave propagation. Important characteristics of elastic waves can be deduced from consideration of slowness curves. Figure~\ref{slowness} schematically shows slowness curves for P-and S-waves for a nearly incompressible DE subjected to a voltage. Note that P-wave does not change its direction, while the S-wave refracts from initial direction. Thus, the P-and S-waves will be split when propagate in such media. It is easy to show that normal to slowness curves has the slope
\begin{equation}\label{ref_ang}
\text{tan}~\theta_{p,s}=\frac{s_{p,s}~\text{sin} \varphi-(d s_{p,s}/d \varphi)~\text{cos} \varphi}{s_{p,s}~\text{cos} \varphi+(d s_{p,s}/d \varphi)~\text{sin} \varphi},
\end{equation}
where $\varphi$ is incident angle and $s_{p,s}=1/c_{p,s}$. Consequently, divergence angle between P-and S-wave can be calculated as $\Delta\theta=\theta_p-\theta_s$.

\section{Results}
\subsection{Ideal dielectric elastomer model}
 First, we consider an ideal dielectric elastomer model~\cite{PhysRevB.76.134113}, namely $\gamma_0=\gamma_2=0$ and $\gamma_1=1$ in~\eqref{genEF}. For an elastic part of the energy function~\eqref{genEF}, here and thereafter, we utilize a model of a compressible neo-Hookean material~\cite{ogden97book}
\begin{equation}\label{neoH}
	\psi_{elas}(\Fder)= \frac \mu{2} (I_1-3)-\mu\ln~J+\left( \frac{K}{2}-\frac{\mu}{3}\right)(J-1)^2, 
\end{equation}
where $I_1=\text{tr}~\Cc$ is the first invariant of the right Cauchy-Green tensor, $\mu$ is the shear modulus and $K$ is the bulk modulus.

For the ideal DE generalized acoustic tensor takes the form
\begin{equation}\label{acten}
\A=a_1^*\n\otimes\n + a_2^* \hat \I,
\end{equation}
where $\n\otimes\n$ is the projection on the direction $\n$ and $\hat \I$ is given in~\eqref{Pn},
\begin{equation}
a_1^*=(K-2\mu/3)J+\mu J^{-1}(1+\n\cdot\B\cdot\n)
\end{equation}
\begin{equation}
\text{and} \quad a_2^*=\mu J^{-1}(\n\cdot\B\cdot\n).
\end{equation}
Consequently, there always exist one pressure and two shear waves for any direction of propagation $\n$, finite deformation $\Fder$ and electric displacement $\D$. Phase velocities of these waves can be calculated as
\begin{equation}\label{phvelIDE}
c_{p}=\sqrt{a_1^* J/\rho_0}\quad \text{and}\quad c_{s}=\sqrt{a_2^* J/\rho_0},
\end{equation}
 where $\rho_0$ is the initial density of the material. Remarkably, while the electroelastic moduli~\eqref{EMT} explicitly depend on electric field, phase velocities of elastic waves do not explicitly depend on the electric field. The dependence of the velocities on electric field is introduced through the electrostatically induced deformation.
Specifically, dielectric elastomer coated with flexible electrodes contracts if a voltage is applied (see Fig.~\ref{deformDE}). Here we consider a case when electric field is orthogonal to the DE layer, i.e. $\D_0=D\sqrt{\mu\epsilon}\ee_2$, and DE layer can freely expands in plane $\ee_1-\ee_3$ (Fig.~\ref{deformDE}). Thus, deformation gradient can be expressed as
\begin{equation}
 \Fder=\lambda(D)\ee_2\otimes\ee_2+\tilde{\lambda}(D)(\I-\ee_2\otimes\ee_2).\label{ut}
 \end{equation}
For an incompressible ideal DE~\cite{PhysRevB.76.134113} $\lambda=(1+D^2)^{-1/3}$ and $\tilde{\lambda}=\lambda^{-1/2}$. It should be noted that these relations approximately  hold even for very compressible materials with $K/\mu\sim 1$ for a range of deformations and electric fields considered here. Recall that neo-Hookean DE is stable for $D\leqslant\sqrt{3}$~\cite{PhysRevB.76.134113} and it will thin down without limit with further increase in electric field. Note that in the absence of an electric field, relations~\eqref{phvelIDE} reduce to $c_{p}=\sqrt{(K+4\mu/3)/\rho_0}$ and $c_{s}=\sqrt{\mu/\rho_0}$.

 For the nearly incompressible DE, phase velocities of P- and S-waves can be calculated as
 \begin{equation}\label{phvelnH_p}
 c_{p}=\sqrt{\left((K + \mu/3+\frac{\mu\left(1+D^2 \text{cos}^2~\varphi\right)}{\left(1+D^2\right)^{2/3}}\right)/\rho_0}
\end{equation}
and
\begin{equation}\label{phvelnH_s}
 c_{s}=\sqrt{\left(1 + D^2 \text{cos}^2~\varphi\right)\left(1+D^2\right)^{-2/3}\mu/\rho_0},
 \end{equation}
 where the propagation direction is defined as $\n=(\text{cos}\varphi,\text{sin}\varphi,0)$.
 Hence, refraction angles of P-and S-waves are
\begin{equation}\label{ref_p}
\text{tan}~\theta_p=\frac{\left(K+\mu/3+\mu\left(1+D^2\right)^{-2/3}\right)\text{tan}~\varphi}{K+\mu/3+\mu\left(1+D^2\right)^{1/3}}
\end{equation}
and
\begin{equation}\label{ref_s}
\text{tan}~\theta_s=\frac{\text{tan}~\varphi}{1+D^2}.
\end{equation}
Remarkably, relations~\eqref{phvelnH_p}, \eqref{phvelnH_s},\eqref{ref_p} and~\eqref{ref_s} approximately hold even for very compressible materials with $K/\mu\sim 1$ for a range of deformations and electric fields considered here.

\begin{figure}[h]
   \centering{\includegraphics[scale=0.35]{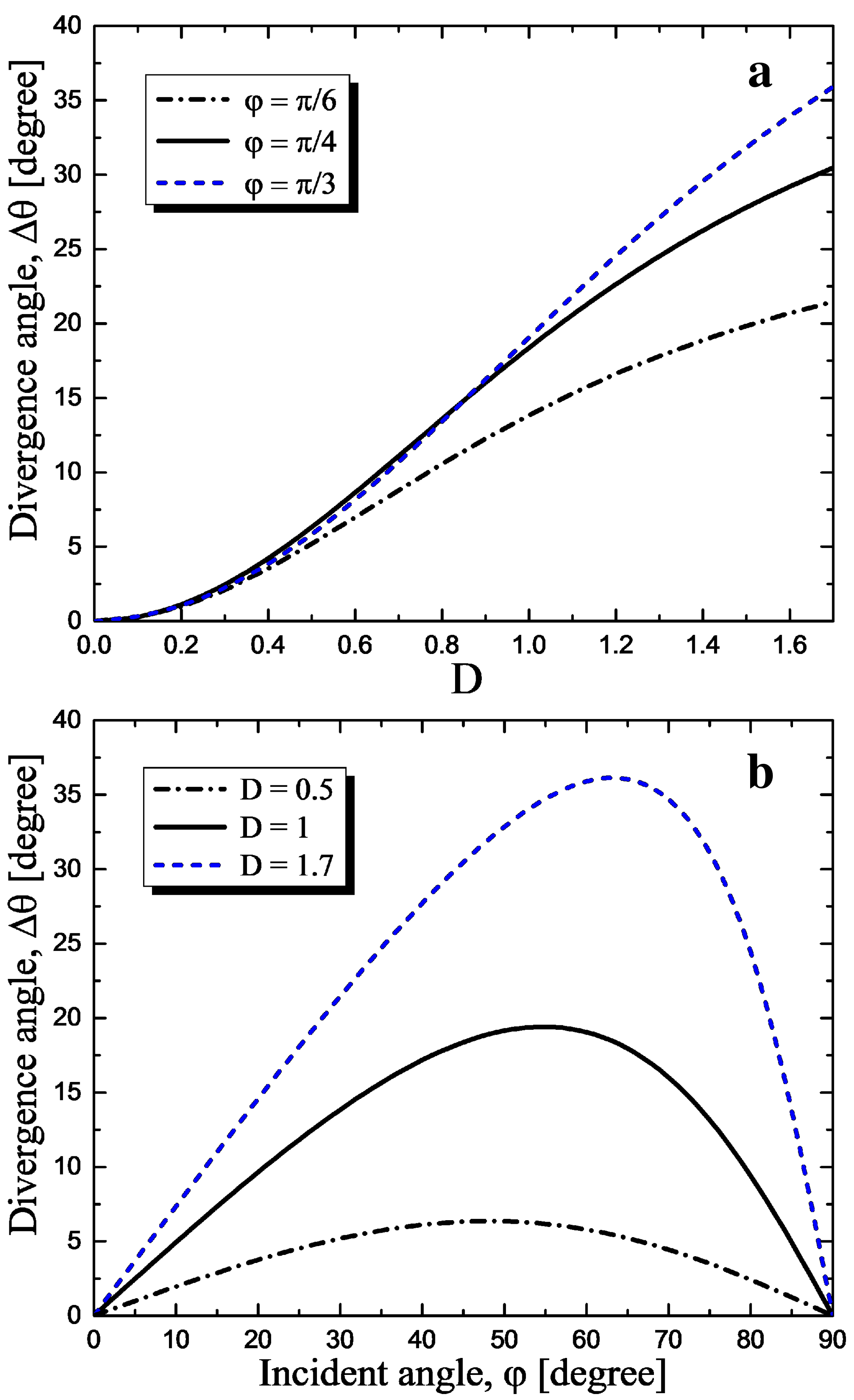}}
  \caption{\footnotesize Divergence angle as functions of non-dimensional electric displacement -- (a), and incident angle -- (b) in the nearly incompressible neo-Hookean DE with $K/\mu=300$.}\label{div_ang}
\end{figure}

 Figure~\ref{div_ang}(a) shows that divergence angle $\Delta\theta$ significantly depends on the value of electric field. In particular, divergence angle monotonically increases with electric field until the limiting value of electric field is achieved ($D=\sqrt{3}$).  Fig.~\ref{div_ang}(b) shows the dependence of the divergence angle $\Delta\theta$ on the incident angle $\varphi$. Clearly, the divergence angle $\Delta\theta$ has a maximum for a certain incident angle $\varphi_0$ depending on the applied electric field. The incident angle $\varphi_0$ corresponding to the maximal divergence angle is given by
\begin{equation}\label{fi0}
\text{tan}~\varphi_0=\sqrt{\frac{\alpha\left(\alpha\mu^3\left(\alpha+27\right)+3 D^2 \alpha_1 \mu \left(\alpha_1-3\mu\right)+ \zeta\right)}{\mu^3\left(\alpha^2+27\right)+ \zeta}},
\end{equation}
where $\alpha=1+D^2$, $\alpha_1=\alpha^{2/3}\left(3 K +\mu\right)$ and $\zeta=9 K \alpha^2\left(\mu^2 + 3K\left(K+\mu\right)\right)$.
The angle $\varphi_0$ monotonically increases with increase in electric field. It is worth mentioning that for $\mu/K\ll 1$,  $\varphi_0=\text{arccos}\left(2+D^2\right)^{-1/2}$.

\begin{figure}[h]
   \centering{\includegraphics[scale=0.55]{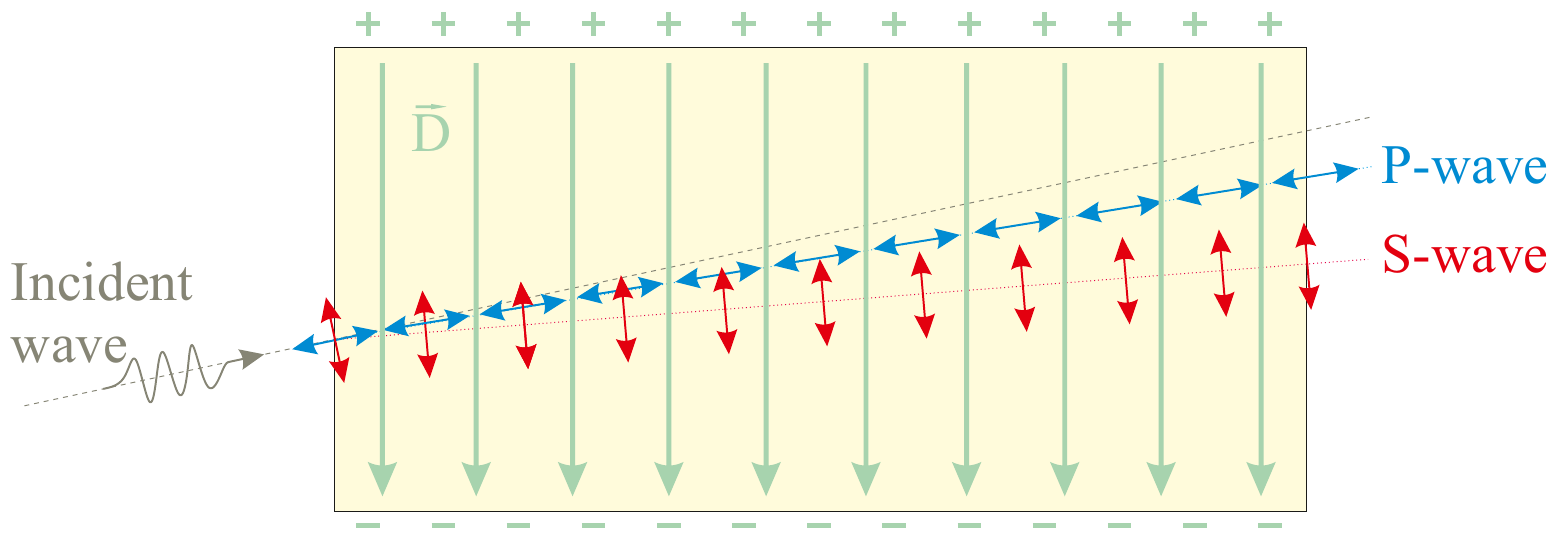}}
  \caption{\footnotesize Schematics of the splitting of P- and S-waves in the compressible neo-Hookean DE. P-and S-wave unidirectionally refract from the initial direction of propagation, but the refraction angles are different. }\label{splcomp}
\end{figure}

\begin{figure}[h]
   \centering{\includegraphics[scale=0.3]{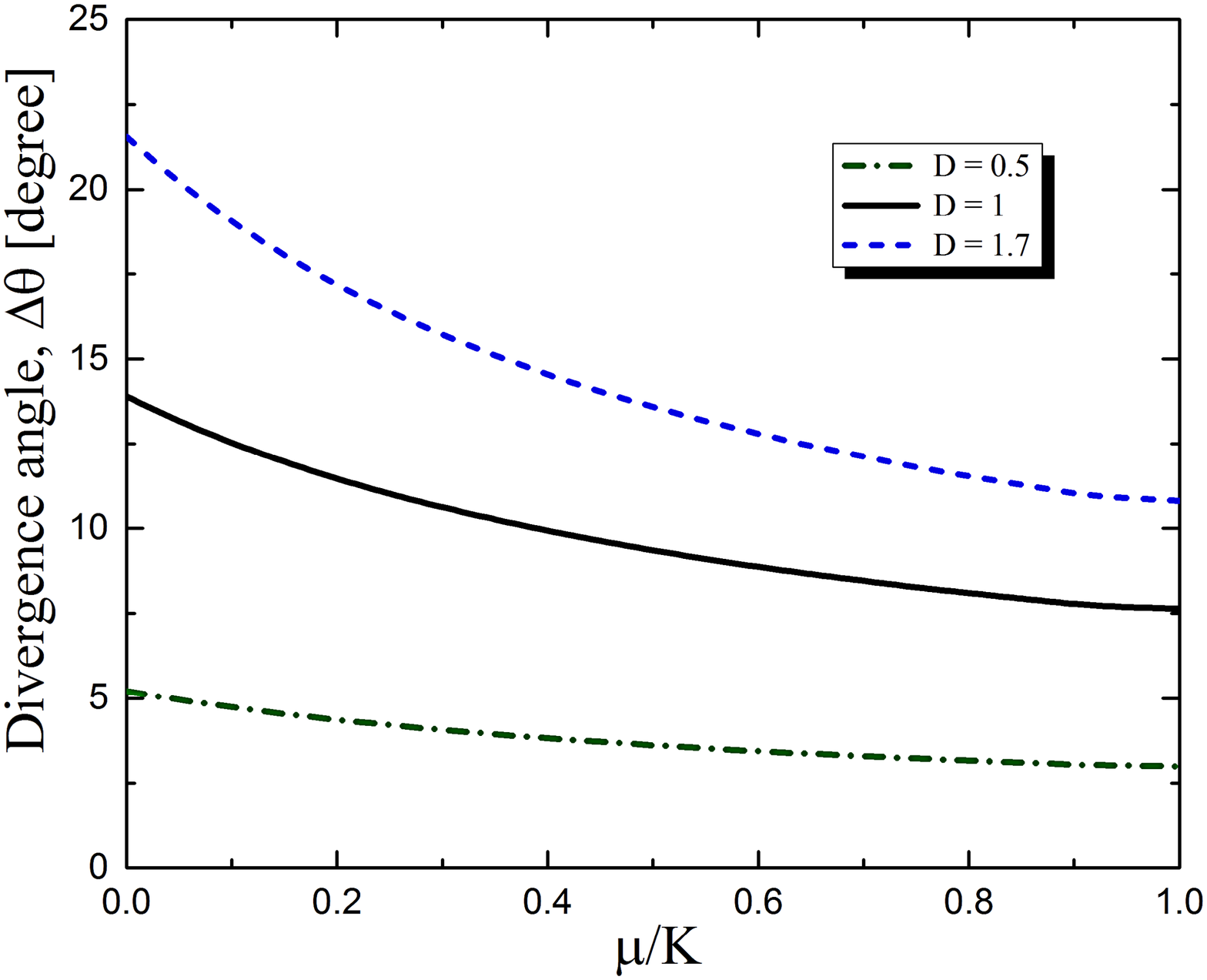}}
  \caption{\footnotesize Divergence angle as functions of material compressibility for the ideal neo-Hookean DE, $\varphi=\pi/6$.}\label{div_vs_com}
\end{figure}

Next we consider the influence of material compressibility on the elastic wave propagation in DEs. It was recently shown for purely mechanical materials that for highly compressible material phase velocity of longitudinal wave significantly depends on the direction of wave propagation and applied deformation~\cite{galich&rudykh2015}. Here, we also observe a strong dependence of P-wave velocity on the direction of wave propagation and deformation induced by an electric field. The more the material compressibility parameter $\mu/K$, the more the influence of electrostatically induced deformation on the elastic waves. In particular, slowness curves of P-wave evolve from circle to an ellipse shape with an increase in electric field. Thus, refraction angle of P-wave differs from the incident angle, namely P-wave refracts along with S-wave. The schematics of the phenomena is illustrated in Fig.~\ref{splcomp}.
It is clear that an increase in compressibility of the DE weakens the disentangling phenomenon.
The dependence of the divergence angles as a function of compressibility for different values of applied electric field is presented in Fig.~\ref{div_vs_com}.

\subsection{Enriched electroactive material model}
Motivated by experimental data~\cite{Wissler&Mazza2007,Li&al2011b}, in this section we investigate elastic wave propagation in DEs described by an enriched electroactive material model~\eqref{genEF}. The model is similar to that considered by Gei et al~\cite{Gei&al2014ijss}. To allow investigation of pressure waves, we extended the model to capture the compressibility effects. Table~\ref{param} summarizes the parameters of the material model~\eqref{genEF} for different experimental data~\cite{Wissler&Mazza2007,Li&al2011b}.
\begin{table}[h]
\caption{Material constants of DE model~\eqref{genEF}.}
\label{param}
\begin{tabular}{c lll}
\hline
    {\footnotesize Reference} & {\footnotesize $\gamma_0$} & {\footnotesize $\gamma_1$} & {\footnotesize $\gamma_2$} \\
\hline
   {\footnotesize Ideal DE~\cite{PhysRevB.76.134113}} & {\footnotesize 0} & {\footnotesize 1} & {\footnotesize 0}  \\
   {\footnotesize Wissler and Mazza~\cite{Wissler&Mazza2007}} & {\footnotesize 0.00104} & {\footnotesize 1.14904} & {\footnotesize $-$0.15008} \\
   {\footnotesize Li et al~\cite{Li&al2011b}} & {\footnotesize 0.00458} & {\footnotesize 1.3298} & {\footnotesize $-$0.33438} \\
\hline
\end{tabular}
\end{table}

\begin{figure}[h]
  \centering{\includegraphics[scale=0.35]{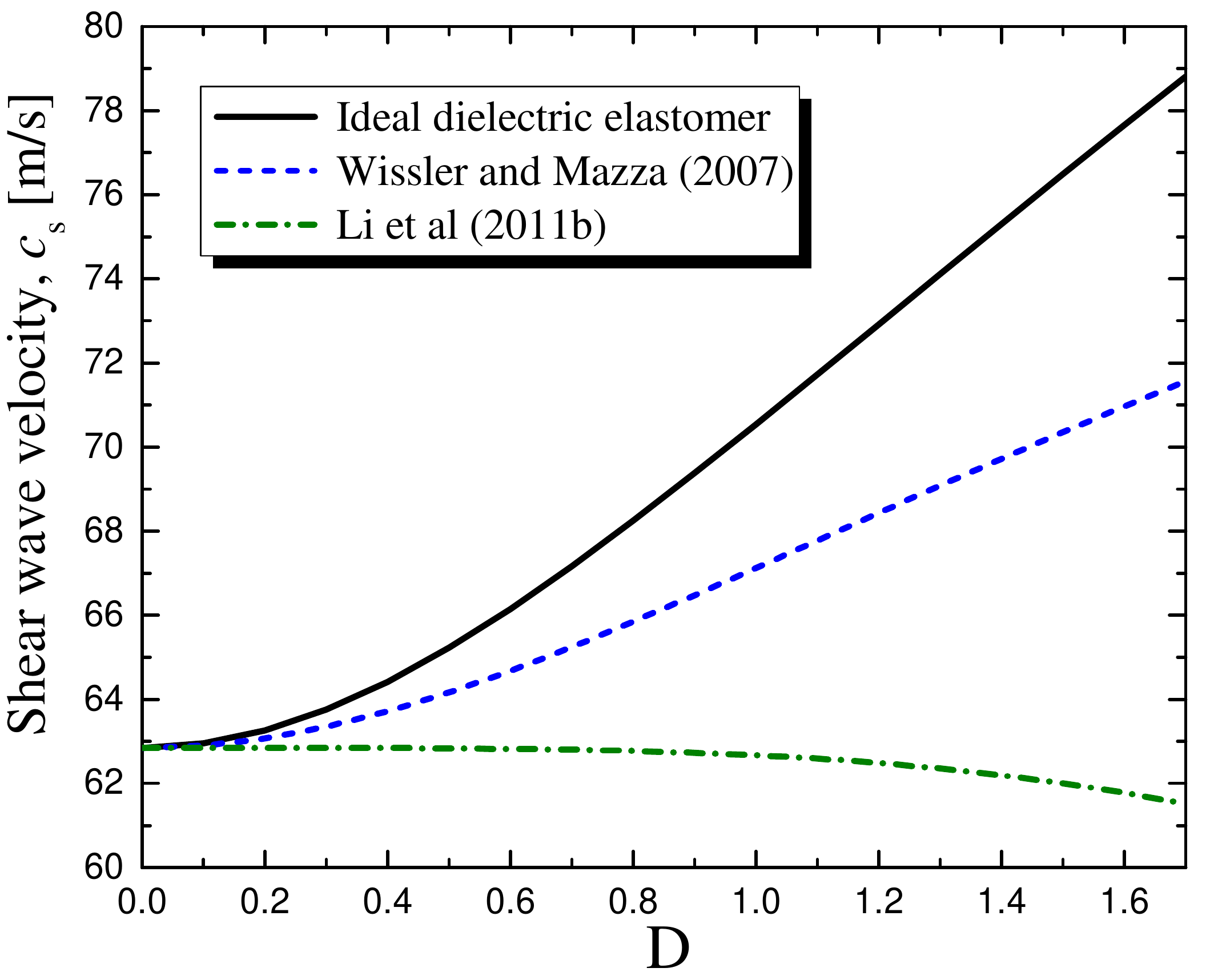}}
 \caption{\footnotesize Phase velocity of in-plane shear wave as functions of non-dimentional electric displacement for $\varphi=0$.}\label{swD}
\end{figure}

\begin{figure}[h]
   \centering{\includegraphics[scale=0.35]{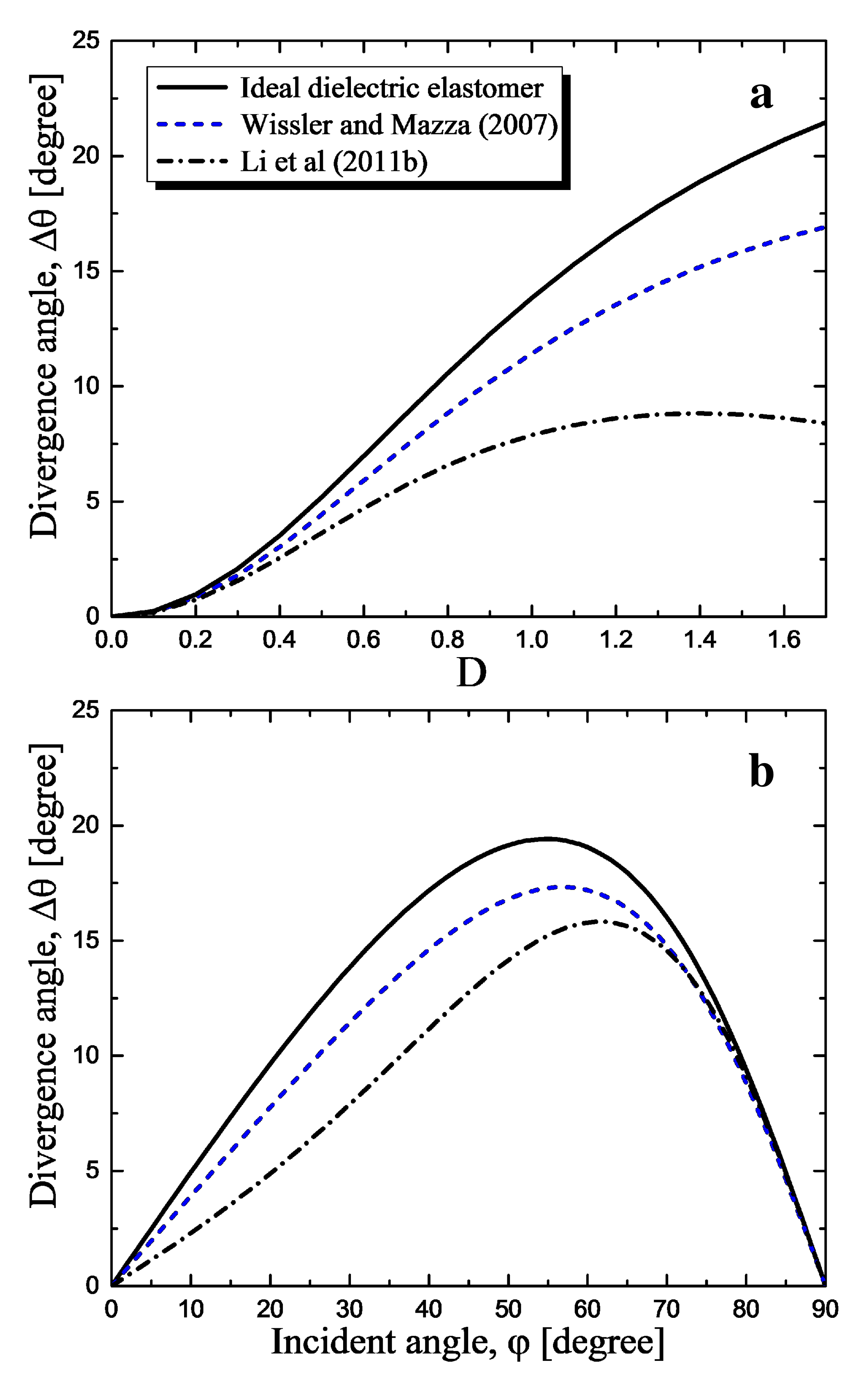}}
  \caption{\footnotesize Divergence angle as functions of non-dimensional electric displacement for $\varphi=\pi/6$ (a) and incident angle for $D=1$ (b) in the nearly incompressible neo-Hookean DE with $K/\mu=300$.}\label{div_ang_combined}
\end{figure}

In case when $\gamma_0\not=0$, $\gamma_2\not=0$, deformation gradient is identical to~\eqref{ut}, and  $\n=\ee_1$, phase velocities takes the following form
\begin{equation}\label{vel_D1}
c_{p}^{(1)}=\sqrt{\left((K-2\mu/3)\lambda^2\tilde{\lambda}^4+\left(1+\tilde{\lambda}^2\right)\mu\right)/\rho_0},
\end{equation}
\begin{equation}\label{vel_D2}
c_{s}^{(2)}=\sqrt{\left(\gamma_2 D^2\lambda+\tilde{\lambda}^2\right)\mu/\rho_0}
\end{equation}
and
\begin{equation}\label{vel_D3}
c_{s}^{(3)}=\tilde{\lambda}\sqrt{\mu/\rho_0}.
\end{equation}
For incompressible materials $\tilde{\lambda}=\lambda^{-1/2}$, and $\lambda$ can be determined from the a solution of the following polynomial equation
\begin{equation}
\lambda^3\left(1+D^2\left(\gamma_1+2\gamma_2\lambda^2\right)\right)=1.
\end{equation}
Hence, phase velocity of the shear wave with polarization  $\vec{g}_2=\ee_2$ (in-plane shear wave) explicitly depends on the electric field in contrast to the ideal DE. Figure~\ref{swD} shows the phase velocity of the in-plane shear wave as a function of the non-dimensional electric field for the ideal DE and enriched DE models. One can see that consideration of the $I_{4e}$ and $I_{6e}$ invariants strongly affects the in-plane shear wave velocity. In particular, the increase in phase velocity of the in-plane shear wave is less prominent (dashed blue curve in Fig.~\ref{swD}) as compared to the ideal DE model result; moreover, the velocity may decrease with an increase in electric field for the particular material (dot-dashed green curve), while for the ideal DE model the velocity always increases.
 These differences indeed affects the splitting mechanism and, thus, different results are produced by these material models.
 Figure~\ref{div_ang_combined} illustrates the difference in the divergence angels for the ideal DE and enriched DE models. In particular, Fig.~\ref{div_ang_combined}(a) shows that divergence angle $\Delta\theta$ is smaller indeed than it was predicted by model of the ideal DE. Moreover, for Li et al. experimental data, divergence angle $\Delta\theta$ decreases with an increase in electric field when certain level of voltage is achieved. Fig.~\ref{div_ang_combined}(b) shows that divergence angle $\Delta\theta$ has a maximum for a certain incident angle $\varphi_1$. Moreover, the incident angle $\varphi_1$, producing the maximum divergence angle, varies for different materials.

Although in this work we perform the fully electromechanical coupling analysis, it is possible to consider only purely mechanical analysis of wave propagation in non-linear electroelastic solid. Thus, contribution of the electroelastic moduli tensors $\M$ and $\K$ is neglected, and only terms of the tensor $\C$ contribute to the generalized acoustic tensor
 Remarkably, this approximation yields a very close results for the phase velocity of pressure wave, and exactly the same relations for the phase velocities of the shear waves. For example, for the deformation gradient~\eqref{ut},  $\n=\ee_1$ and  $\D_0=D\sqrt{\mu\epsilon}\ee_2$ {\it classic} acoustic tensor takes the form
\begin{equation}\label{approx}
\begin{split}
\Q=&\ \Q_{elas}+\mu D^2(\gamma_0 J^{-2}+\gamma_1\tilde{\lambda}^{-4}+\gamma_2\lambda^2\tilde{\lambda}^{-4})\ee_1\otimes\ee_1\\
&+\mu \gamma_2 D^2 \tilde{\lambda}^{-2}\ee_2\otimes\ee_2.
\end{split}
\end{equation}
It is easy to see that the resulting expressions for phase velocities of S-waves are the same as expressions~\eqref{vel_D2} and~\eqref{vel_D3}, and expression for the phase velocity of P-wave has the form
\begin{equation}\label{vel_D_nc}
\begin{split}
c_{p}^{(1)}=&\ \sqrt{\mu/\rho_0}\Bigl(1+\tilde{\lambda}^2+(K/\mu-2/3)\lambda^2\tilde{\lambda}^4\\
&+D^2(\gamma_0\lambda^{-2} + \gamma_1 +\gamma_2\lambda^2)\tilde{\lambda}^{-4}\Bigr)^{-1/2}.
\end{split}
\end{equation}
Expression~\eqref{vel_D_nc} for P-wave velocity differs from~\eqref{vel_D1} in the additional term containing electric displacement magnitude. However, according to our observations for $K/\mu\gtrsim 100$, this term can be neglected for small and moderate levels of electric displacement.
\section{Concluding remarks}
In this paper, we considered P-and S-elastic waves in soft DEs subjected to an electric field. To allow consideration of P-wave propagation, we utilized the ideal and enriched material models containing {\it compressibility} effects. Explicit relations of the generalized acoustic tensor and phase velocities of elastic waves for the DE subjected to an electric field were presented.   We found that, for the ideal DE elastic wave propagation is explicitly independent of the applied electric field, and it is influenced through deformation induced via bias electric field; this is despite the fact that all electroelastic moduli tensors explicitly depend on the applied electric field. However, for the enriched material model, including all three electroelastic invariants $I_{4e},I_{5e}$ and $I_{6e}$, elastic wave propagation explicitly depends on the applied electric field; namely, phase velocity of in-plane shear wave decreases when electric field is applied.

These findings were applied to explore the phenomenon of disentangling of P-and S-elastic waves in DEs by an electric field. We showed that divergence angle between P-and S-wave strongly depends on the value of the applied electric field and direction of wave propagation. Moreover, we found that an increase in the material compressibility weakens separation of P-and S-waves. This is due to the fact that  P-wave also refracts along with the S-wave from the initial direction of wave propagation, while for nearly incompressible materials the change in its direction is negligible. The phenomenon can be used to manipulate elastic waves by a bias electrostatic field; this can be beneficial for an application at the small length-scale devices, such as micro-electromechanical systems, where electric field is a preferred control parameter.

\section*{Acknowledgements}
This research was supported by the ISRAEL SCIENCE FOUNDATION (grant 1550/15). SR gratefully acknowledges the support of Taub Foundation through the Horev Fellowship --  Leaders in Science and Technology.

\appendix
\section{\\Notation of a tensor isomer} \label{Appendix A}
 Here we follow the notation of an isomer firstly introduced by Ryzhak~\cite{ryzhak93jmps,nikitin&ryzhak2008}.
 Let $\mathcal{S}$ be a third-rank tensor with the following representation as the sum of a certain number of triads:
\begin{equation}
\mathcal{S}=\ab_1\otimes\ab_2\otimes\ab_3+\bb_1\otimes\bb_2\otimes\bb_3+\ldots
\end{equation}
Let (ijk) be some permutation of the numbers (123). Then the isomer $\mathcal{S}^{(ijk)}$ is defined to be the third-rank tensor determined by the relation
\begin{equation}
 \mathcal{S}^{(ijk)}=\ab_i\otimes\ab_j\otimes\ab_k+\bb_i\otimes\bb_j\otimes\bb_k+\ldots
\end{equation}
Analogously, we can define isomers for higher-rank tensors.

\section{\\Components of the electroelastic moduli and generalized acoustic tensors} \label{Appendix B}
Here we introduce the following notations
\begin{equation}
\begin{split}
L_{ab}=&\vec{a}\cdot\mathbf{L}\cdot\vec{b},L_{ab}^{(2)}=\vec{a}\cdot\mathbf{L}^2\cdot\vec{b},\mathbf{L}^2=\mathbf{L}\cdot\mathbf{L}, \\
L_{ab}^{(-1)}=&\vec{a}\cdot\mathbf{L}^{-1}\cdot\vec{b},L_{ab}^{(-2)}=\vec{a}\cdot\mathbf{L}^{-2}\cdot\vec{b},\mathbf{L}^{-2}=\mathbf{L}^{-1}\cdot\mathbf{L}^{-1}, \\
L_{ab}^{(3)}=&\vec{a}\cdot\mathbf{L}^{3}\cdot\vec{b},\mathbf{L}^{3}=\mathbf{L}^2\cdot\mathbf{L}, L_{ab}^{(-3)}=\vec{a}\cdot\mathbf{L}^{-3}\cdot\vec{b},\\
\mathbf{L}^{-3}=&\mathbf{L}^{-2}\cdot\mathbf{L}^{-1},\vec{a}\otimes\vec{b}^s=\frac{1}{2}\left(\vec{a}\otimes\vec{b}+\vec{b}\otimes\vec{a}\right),\\ a_b=&\vec{a}\cdot\vec{b},a_b^2=(a_b)^2,
\end{split}
\end{equation}
where $\mathbf{L}$ is an arbitrary second-rank tensor; $\vec{a}$ and $\vec{b}$ are arbitrary vectors.

\begin{equation}
\begin{split}
\C_{4e}=&\ B_{DD}^{-1}(\I\otimes\I+\I\otimes\I^{(1342)}),\\
\C_{5e}=&\ 2(\D\otimes\I\otimes\D^{(2134)}-\D\otimes\D\otimes\I-\I\otimes\D\otimes\D)\\
&+D_D (\I\otimes\I+\I\otimes\I^{(1342)}),\\
\C_{6e}=&\ 2(\D\otimes\I\otimes(\B\cdot\D)^{(2134)}+\B\otimes\D\otimes\D^{(1432)}\\
&+\D\otimes\B\otimes\D^{(2134)}+(\B\cdot\D)\otimes\I\otimes\D^{(2134)}\\
&+\D\otimes\D\otimes\B^{(1324)}+\D\otimes\B\otimes\D\\
&-\I\otimes(\B\cdot\D)\otimes\D-\I\otimes\D\otimes(\B\cdot\D)\\
&-\D\otimes(\B\cdot\D)\otimes\I-(\B\cdot\D)\otimes\D\otimes\I)\\
&+B_{DD}(\I\otimes\I+\I\otimes\I^{(1342)}).
\end{split}
\end{equation}

\begin{equation}
\begin{split}
\Q_{4e}=&\ 2\ B_{DD}^{(-1)}\n\otimes\n,\\
\Q_{5e}=&\ 2((D_D+D_n^2)\n\otimes\n+D_n^2\hat{\I}-2D_n\D\otimes\n^{s})),\\
\Q_{6e}=&\ 2((2B_{Dn}D_n+B_{DD})\n\otimes\n+2B_{Dn}D_n\hat{\I}\\
&-2B_{Dn}\D\otimes\n^{s}+D_n^2\B+2D_n(\B\cdot\n)\otimes\D^s\\
&-2D_n(\B\cdot\D)\otimes\n^s+B_{nn}\D\otimes\D).
\end{split}
\end{equation}

\begin{equation}
\begin{split}
\M_{4e}=&\ -2\I\otimes(\B^{-1}\cdot\D),\\
\M_{5e}=&\ 2(\D\otimes\I+\D\otimes\I^{(213)}-\I\otimes\D),\\
\M_{6e}=&\ 2(\D\otimes\B+\D\otimes\B^{(213)}+(\B\cdot\D)\otimes\I\\
&+(\B\cdot\D)\otimes\I^{(213)}-\I\otimes(\B\cdot\D)).
\end{split}
\end{equation}

\begin{equation}
\begin{split}
\R_{4e}=&\ -2\n\otimes(\B^{-1}\cdot\D),\\
\R_{5e}=&\ 2(D_n\I+\D\otimes\n-\n\otimes\D),\\
\R_{6e}=&\ 2(D_n\B+\D\otimes(\B\cdot\n)+B_{Dn}\I\\
&+(\B\cdot\D)\otimes\n-\n\otimes(\B\cdot\D)).
\end{split}
\end{equation}

\begin{equation}
\begin{split}
\eta=&\gamma_0^2\left(2\BtnnInv-\B^{-1}:\B^{-1}-(\BnnInv)^2\right)+ \gamma_2^2\left(2\Btnn-\B:\B-(\Bnn)^2\right)\\
&+\left(\gamma_0\left(\I:\B^{-1}-\BnnInv\right)+\gamma_1\right)^2+\left(\gamma_2\beta+\gamma_1\right)^2+2\gamma_0\gamma_2\left((\I:\B^{-1})\beta-I_1\BnnInv-1\right)
\end{split}
\end{equation}

\begin{equation}
\begin{split}
\A_{e}=&\  a_1 \n\otimes\n+a_2 \I +a_3 \D\otimes\D+a_4 \n\otimes\D^s+a_5 \B + a_6 (\B\cdot\n)\otimes\D^s\\
&+ a_7 (\B\cdot\D)\otimes\n^s+a_8 (\B\cdot\n)\otimes\n^s+a_9 \B^2+ a_{10} (\B^2\cdot\n)\otimes\D^s\\
&+a_{11} (\B^2\cdot\D)\otimes\n^s+a_{12} (\B^2\cdot\n)\otimes\n^s+a_{13} (\B\cdot\n)\otimes(\B\cdot\n)\\
&+ a_{14} \B^3+ a_{15} (\B^3\cdot\n)\otimes\D^s+a_{16} (\B^3\cdot\D)\otimes\n^s\\
& +a_{17} (\B^2\cdot\n)\otimes(\B\cdot\n)^s+a_{18} \B^{-1}+a_{19} (\B^{-1}\cdot\n)\otimes\n^s\\
&+a_{20} (\B^{-1}\cdot\n)\otimes\D^s+a_{21} (\B^{-1}\cdot\D)\otimes\n^s\\
&+a_{22} (\B^{-1}\cdot\n)\otimes(\B\cdot\n)^s+a_{23} (\B^{-2}\cdot\D)\otimes\n^s,
\end{split}
\end{equation}
where
\begin{equation}
\begin{split}
a_1=& \gamma_0^3\left(\left(\BtddInv-(\BndInv)^2\right)\I:\B^{-1}-\BnnInv\BtddInv-\BtrddInv+2\BndInv\BtndInv\right)+\gamma_1^3 D_D\\
&+\gamma_2^3\left(\beta\Btdd-\Btrdd\right)+2\gamma_0\gamma_1\gamma_2\left(\beta\BddInv+\Bnd\BndInv+\beta_4\Bdd-D_D\right)\\
&+\gamma_0\gamma_1^2\left(\beta_4 D_D+\BddInv\right)+\gamma_0\gamma_2^2\left(\beta_4\Btdd-3\Bdd+2\beta D_D+2\BndInv\Btnd\right)\\
&+\gamma_1\gamma_0^2\left(2\beta_4\BddInv-\BtddInv+(\BndInv)^2\right)+\gamma_1\gamma_2^2\left(2\beta\Bdd-\Btdd\right)\\
&+\gamma_2\gamma_0^2\left(\beta\BtddInv-3\BddInv+2\beta_4 D_D+\BndInv\left(2 D_n-I_1\BndInv\right)\right)+\gamma_2\gamma_1^2\left(\beta D_D +\Bdd\right),\\
a_2=&\gamma_1^3 D_n^2+\gamma_2^3\beta (\Bnd)^2+2\gamma_0\gamma_1\gamma_2 D_n \left(\beta_4\Bnd-D_n\right)+\gamma_0\gamma_1^2\beta_4 D_n^2\\
&+\gamma_0\gamma_2^2\left(\beta_4\Bnd-2 D_n\right)\Bnd+\gamma_1\gamma_2^2\Bnd\left(\Bnd+2\beta D_n\right)+\gamma_2\gamma_1^2\left(\beta D_n+2\Bnd\right)D_n,\\
a_3=& \gamma_2^2\Bigr(\gamma_2\left(I_1\left(\Btnn-(\Bnn)^2\right)+\Bnn\Btnn-\Btrnn\right)+\gamma_0\left(\Bnn-\BnnInv\Btnn+(\I:\B^{-1})\left(\Btnn-(\Bnn)^2\right)\right)\\
&+\gamma_1\left(\Btnn-(\Bnn)^2\right)\Bigl),\\
a_4=& 2\Bigr(\gamma_0\gamma_1\gamma_2\left(D_n-2\beta_4\Bnd\right)-\gamma_1^3 D_n-\gamma_2^3\left(I_1\Btnd-\Btrnd\right)-\gamma_0\gamma_1^2\beta_4 D_n\\
&+\gamma_0\gamma_2^2\left(3\Bnd-\beta_4\Btnd+I_1\left(\BndInv\Bnn-2 D_n\right)+D_n\Bnn-\BndInv\Btnn\right)+\gamma_1\gamma_2^2\Bnd\left(\Bnn-2 I_1\right)\\
&+\gamma_2\gamma_0^2\left(\Bnn\left(\BndInv(\I:\B^{-1})-\BtndInv\right)-2 D_n\beta_4\right)-\gamma_2\gamma_1^2\left(\beta D_n+2\Bnd\right)\Bigl),\\
a_5=&\gamma_2\Bigr(\gamma_2^2\Bnd\left(2 D_n\beta-\Bnd\right)+2\gamma_0\gamma_1\beta_4 D_n^2+\gamma_0\gamma_2 D_n\left(2\beta_4\Bnd-D_n\right)\\
&+2\gamma_1\gamma_2\beta D_n^2+\gamma_1^2 D_n^2\Bigl),\\
a_6=& 2\gamma_2\Bigr(\gamma_2^2\left(D_n\Btnn+I_1\left(\Bnd-D_n\Bnn\right)\right)+\gamma_0\gamma_1\beta_4 D_n+\gamma_0\gamma_2\left((\I:\B^{-1})\left(\Bnd-D_n\Bnn\right)-\BnnInv\Bnd\right)\\
&+\gamma_1\gamma_2\left(\Bnd+\beta D_n\right)+\gamma_1^2 D_n\Bigl),\\
a_7=& -2\gamma_2\Bigr(\gamma_2^2\beta\Bnd+2\gamma_0\gamma_1\beta_4 D_n+\gamma_0\gamma_2\left(\beta_4\Bnd-2 D_n\right)+2\gamma_1\gamma_2\beta D_n+\gamma_1^2 D_n\Bigl),\\
a_8=& 2\gamma_2\Bigr(\gamma_0\gamma_2\left(\BndInv\left(I_1 D_n-\Bnd\right)-D_n^2\right)+D_n \Bigr(\gamma_0^2 \left(\BndInv(\I:\B^{-1})-\BtndInv\right)-\gamma_2^2 \Btnd\\
&-\gamma_0\gamma_1 \BndInv-\gamma_1\gamma_2 \Bnd\Bigl)\Bigl),\\
a_9=&\gamma_2^2 D_n\left(\gamma_2\left(\beta D_n-2\Bnd\right)+D_n\left(\gamma_0\beta_4-\gamma_1\right)\right),a_{10}= 2\gamma_2^2\left(\gamma_2\left(I_1 D_n-\Bnd\right)+\gamma_0\beta_4 D_n\right),\\
a_{11}=& 2\gamma_2^2\left(\gamma_2\left(\Bnd-\beta D_n\right)-D_n\left(\gamma_0\beta_4-\gamma_1\right)\right),a_{12}=-2\gamma_0\gamma_2^2 D_n \BndInv,\\
a_{13}=&\gamma_2^2 D_n\left(\gamma_2\left(2\Bnd-I_1 D_n\right)+\gamma_1 D_n-\gamma_0 D_n (\I:\B^{-1})\right),\\
a_{14}=&-\gamma_2^3 D_n^2, a_{15}=-2\gamma_2^3 D_n, a_{16}=-a_{15},a_{17}=-2 a_{14},\\
a_{18}=&-\gamma_0\left(\gamma_1 D_n+\gamma_2\Bnd\right)^2, a_{19}=-2\gamma_0^2\BndInv\left(\gamma_1 D_n+\gamma_2\Bnd\right),\\
a_{20}=& 2\gamma_0\gamma_2\Bnn\left(\gamma_1 D_n+\gamma_2\Bnd\right),\\
a_{21}=& 2\gamma_0\left(\gamma_2\gamma_0\left(D_n-\beta_4\Bnd\right)-\gamma_2\beta\left(\gamma_1 D_n+\gamma_2\Bnd\right)-\gamma_1\gamma_0\beta_4 D_n\right) ,\\ 
a_{22}=& 2\gamma_0\gamma_2 D_n\left(\gamma_1 D_n+\gamma_2\Bnd\right),a_{23}= 2\gamma_0^2\left(\gamma_1 D_n+\gamma_2\Bnd\right),\\
 \beta=& I_1-B_{nn}\quad\text{and}\quad \beta_4=\I:\B^{-1}-\BnnInv.
\end{split}
\end{equation}

\section*{References}

\nocite{*}
\bibliographystyle{abbrv}
\bibliography{DE}

\end{document}